\documentclass[12pt]{article}
\usepackage{epsfig}
\usepackage{longtable}

\def\Teff{$T_{\rm eff}$}
\def\logg{$\log\,g$}

\def\Vt{V${\rm t}$}
\begin{document}

\title{
MOLYBDENUM IN THE OPEN CLUSTER STARS}
\author{T.~Mishenina$^{1}$, E.~Shereta$^{1}$, M.~Pignatari$^{2,3,6,7}$, 
  G.~Carraro$^{4}$, \\ T.~Gorbaneva$^{1}$,  C.~Soubiran$^{5}$\\  
$^{1}$Astronomical Observatory,  Odessa National University, \\     
           Marazlievskaia Str., 1V, 65014, Odessa, Ukraine \\ {\tt tmishenina@ukr.net }\\
					$^{2}$	E.A. Milne Centre for Astrophysics,  Department of Physics \\ \& Mathematics, 
University of Hull, HU6 7RX, United Kingdom\\ {\tt mpignatari@gmail.com}\\
$^{3}$	Konkoly Observatory, Hungarian Academy of Sciences, \\
Konkoly Thege Miklos ut 15-17, H-1121 Budapest, Hungary \\
$^{4}$ Dipartimento di Fisica e Astronomia, \\Universita di Padova, I-35122, Padova, Italy\\{\tt giovanni.carraro@unipd.it}\\
 $^{5}$  Laboratoire d'astrophysique de Bordeaux, \\ Universit\'e Bordeaux, CNRS, B18N, all\'ee Geoffroy Saint-Hilaire,\\33615 Pessac, France \\ {\tt caroline.soubiran@u-bordeaux.fr}\\
$^{6}$  Joint Institute for Nuclear Astrophysics\\ - Center for the Evolution of the Elements, USA \\
$^{7}$  NuGrid Collaboration https://nugrid.github.io/}
%\date{\today}
%\end{center}
\maketitle 
\begin{abstract}
Molybdenum abundances in the stars from 13 different open clusters were determined. High-resolution stellar spectra were obtained using the VLT telescope equipped with the UVES spectrograph on Cerro Paranal, Chile. The Mo abundances were derived in the LTE approximation from the Mo I lines at 5506 \AA~ and 5533 \AA.  A comparative analysis of the behaviour of molybdenum in the sampled stars of open clusters and Galactic disc show similar trends of decreasing Mo abundances with increasing metallicities; such a behaviour pattern suggests a common origin of the examined populations. On the other hand, the scatter of Mo abundances in the open cluster stars is slightly greater, 0.14 dex versus 0.11 dex. The results are discussed, considering the abundance trends with the age of clusters and distances from the center of the Galaxy.  
\end{abstract}
\par

%\end{document}
%\tt selena_a@ukr.net, \tt clumpstars@ukr.net}
\section{Introduction}

The distribution of elemental abundances in different Galactic substructures is essential for the study of the  evolution of the Milky Way, and of the nucleosynthesis processes in stars. The nucleosynthesis of Molybdenum (Mo, Z=42) is not been fully understood. Galactic chemical evolution (GCE) simulations estimated that about 40\% of the solar Mo is made by the s-process(slow process of neutron capture), while the r-process (rapid process of neutron capture) contribution is uncertain (e.g., [1]).  Mo is most likely characterized also by a  contribution  from different explosive nucleosynthesis processes in collapsing supernovae (CCSNe) that is not associated to the r-process. Therefore, the study of Mo abundance observed in the Galactic disc is a challenging task for current GCE simulations [2, 3].

From a theoretical point of view (in terms of nucleosynthesis), low- and moderate-mass asymptotic giant branch (AGB) stars (the main s-process) [4, 5, 6, etc.], massive stars (the weak s-process) (e.g., [7, 8, 9, 10]) and fast-rotating massive stars [11] are the different s-process sources of Mo. Neutrino-driven winds from CCSNe may play an important role [12, 13, 14, 15, etc.] , while the r-process contribution might be associated to one or more possible sources, e.g., neutron star -– neutron star  mergers [16, 17, 18, 19, 20]. The studies of the r-process and its components in the Galaxy are actively carried out nowadays (for further details refer to [18, 21, 22]. The underproduction of light isotopes of molybdenum (92 and 94 Mo) in proton-capture reactions in CCSNe (e.g., [23, 24]) remains matter of debate. Additionally, some other processes have been investigated – like the intermediate neutron-capture process (the i-process, [25]) from different stellar sources, e.g, from rapidly-accreting white dwarfs (RAWDs) [26]. 

From an observational point of view, the Mo abundance was thoroughly investigated in metal-poor stars (e.g., [27, 28, 29, 30, 31, 32]). In the studies of turnoff stars, Peterson [28,29] found overabundances of Mo (up to 1 dex) which had not been detected in the field and globular cluster giants (e.g. [33, 34, 35, 36]). In order to explain the obtained overabundances of Mo, Peterson [28] suggested that it was the low-entropy domain of a high-entropy wind (HEW) above the neutron star formed in a Type II supernova that was responsible for the additional molybdenum production (e.g. [37]). 
Later, Hansen et al. [31], having conducted a survey of 71 metal-poor field stars, confirmed the expected enrichment from HEWs and suggested that several other sources – namely, the proton-capture process (p-process), the s-process or the r-process could also be responsible for the Mo formation.

By determining the Mo abundances in the Galactic disc stars, Mishenina et al. [3] expanded the range of metallicities (up to [Fe/H] $\approx$ 0.3 dex) for stars with specified Mo abundances. The analysis  using  two different GCE codes [1, 38] showed that, in comparison with the calculated s-process yields in AGB stars, there was a missing s-process contribution of about  20\% (e.g. [1, 6]). Some missing production from explosive nucleosynthesis was also identified from looking at the overall pattern of [Mo/Fe].  Prantzos et al. [39] also estimated the contributions of the s-process  and r-process to the solar isotopic and elemental abundances:  0.497 from the s-process (about 10\% higher compared to [3]), 0.275 for the r-process and 0.228 from the p-process.

When investigating the Mo enrichment of the Galactic disc with a wider sample of Galactic substructures, we have focused on open clusters (OCs) which belong to the Galactic disc, but at the same time can be considered separate subpopulations of this formation. The study of stellar abundances in OCs  may provide some additional information about the disc enrichment and structure. If we accept the scenario where  open clusters and stellar associations are formed within dense molecular clouds showing  chemical compositions similar to those of their members, and later they are dispersed throughout the disc, contributing to its formation [40, 41], then it is crucial to carry out a comparative study of distinguishing aspects of the chemical composition of stars in OCs and Galactic disc.

This study aims to determine the Mo abundances in the open cluster stars, to examine the molybdenum enrichment of OCs and analyse the features of the Mo abundance behaviour in the Galactic disc.

\section{Stars under examination, observations and stellar parameters}

For the purposes of this study, we have selected stars belonging to 13 open clusters from the list of stars which, along with the detailed description of the spectral data and determinations of main parameters, was taken from [42, 43]. The main parameters of the examined clusters are given in Tabl.\ref{par}, including galactic coordinates (for J2000.0), galacto-centric distance $R_{GC}$ and age, while atmospheric parameters of cluster stars are presented in Tabl.\ref{parmo}, with memberships in the last column (Mem). Now let us give a summary of the main characteristics of spectra, stars and clusters.

The spectra of the selected stars were obtained using a high-resolution ultraviolet and visible light echelle spectrograph (UVES) ensuring the resolving power R = 47,000 at the wavelength range 4760--6840 \AA\AA, which is installed at the VLT telescope operated by the European Southern Observatory (ESO) on Cerro Paranal, Chile. In the study by Mishenina et al. [42], we adopted stellar parameters reported by Carraro et al.[40] and Magrini et al.[44]. Effective temperatures \Teff~ were derived from the photometric data using the calibration described in [45] further adjusted for the correlation between the iron abundance determined from a certain line and lower-level excitation potential for that line. Surface gravities \logg~ were determined using a canonical formula (for further details refer to [40]). 

In order to test the atmospheric parameters reported in [40] and in our work [42], two stars in each of two open clusters, namely Ruprecht 4 and Ruprecht 7, were selected; we compared the measured equivalent widths EWs of lines in these two stars, as well as the effective temperatures \Teff~ estimated in [42] using calibrations of the line-depths ratios for different lower-level excitation potentials developed by Kovtyukh et al. [46]. The comparison yielded a good agreement (see in detail in [42]) for both equivalent widths EWs and effective temperatures \Teff~, and the atmospheric parameters reported in [40, 44] were adopted in the study. In [43], the parameters were determined by applying the afore-described technique, in particular \Teff~ was determined by calibrating the line-depths ratios for different lower-level excitation potentials [46]. The gravity \logg~ was obtained from the iron ionisation balance. The microturbulent velocity \Vt~ was derived factoring in that the iron abundance logA (Fe) derived from the specified Fe I line did not correlate with the EW of that line. A comparison of the atmospheric parameters with those reported in the literature showed a good agreement [43].

\begin{table*}
\caption{Main characteristics of the investigated clusters}
\label{par}
\begin{tabular}{lccccl}
\hline
 Name       &     l   &     b   &$R_{GC} $ &    age  & References \\
           &    deg  &    deg  & kpc &    Gyr  &       \\
\hline
Berkeley 75  & 234.30 &  11.12 & 9.8 & 3.0&  [47]\\
Berkeley 25  &226.60  &  9.69   &11.3 &4.0&  [47]\\
Ruprecht 7   &225.44  &  4.58   & 6.5  &0.8 &   [48]\\
Ruprecht 4   &222.04  & 5.31   & 4.9   &0.8 & [48]\\
Berkeley 73  &215.28  &9.42    &9.7   &1.5 & [47]\\
NGC6192     &340.65   &2.12   &1.5    &0.18 &  [49]\\
NGC6404     &355.66   &-1.18   &1.7 &0.5 &  [50]\\
NGC6583     &9.28      &-1.53   &2.1 &1.0  & [50]\\
Collinder 110 &209.649 &-1.927 &10.2 &1.3 & [51]\\
Collinder 261 &301.684 &-5.528 &7.5 &7.0 & [52]\\
NGC 2477    &253.563 &-5.838 &8.9 &0.6 &  [53\\
NGC 2506    &230.564 &9.935  &10.9 &1.9 &  [54]\\
NGC 5822    &321.577  &3.585 &7.9 &0.45 & [55]\\
\hline
\end{tabular}
\end{table*}

%Table 2. Main parameters of stars in the examined clusters and respective Mo abundances.

%\begin{table*}
\begin{longtable}{lccccccl}
\caption{Main characteristics and Mo abundances}
\label{parmo}
%\begin{tabular}{lcccccccl}
\\\hline
Cluster/ID   & \Teff, K & \logg  & [Fe/H] & \Vt, km/s & [Mo/Fe]& $\sigma$, $\pm$ &Mem\\
\hline  
%gig   &  &  &  &  &  &    &  &  \\
%\endhead
\hline
 Berkeley 25 &  &  &  &  &  &    &    \\
 10	&5000	&2.9	&0.10	&1.65		 &0.09	&	&NM \\
 12	&4870	&2.75	&-0.20	&1.50		 &0.12	&	&M  \\
 13	&4860	&2.65	&-0.17	&1.73		 &0.24	&	&M  \\
 Berkeley 73 &  &  &  &  &  &    &    \\
12	&5030	&2.78	&-0.39	&1.40		 &0.35	&0.22	&NM \\
13	&5730	&4.15	&0.17	&0.99		 &-0.14	&0.16	&NM \\
15	&5070	&3.12	&-0.38	&1.04		 &0.28	&0.04	&NM \\
16	&4890	&2.71	&-0.18	&1.45		 &0.09	&0.05	&M  \\
18	&4940	&2.88	&-0.27	&1.32		 &0.17	&0.03	&M  \\
19	&5870	&4.23	&-0.03	&1.40		 &0.10	&0.07	&NM \\
 Berkeley 75  &  &  &  &  &  &    &    \\
9	&4968	&2.57	&-0.44			&1.5	 &0.06	&0.01	&NM \\
22	&5180	&3.37	&-0.22	&1.21		 &0.13	&0.06	&M  \\
 Collinder 110 &  &  &  &  &  &    &    \\
1122	&4954	&2.60	&-0.06	&1.20		 &-0.03	&0.06	&M  \\
1134	&4940	&2.60	&0.02	&1.20		 &-0.19	&0.03	&M  \\
1149	&4906	&2.60	&-0.01	&1.20		 &-0.03	&0.08	&M  \\
1151	&4956	&2.60	&0.02	&1.20		 &-0.04	&0.02	&M  \\
2129	&4933	&2.60	&-0.04	&1.20		 &-0.08	&0.00	&M  \\
3122	&4758	&2.40	&-0.03	&1.00		 &-0.20	&0.03	&M  \\
 Collinder 261 &  &  &  &  &  &    &    \\ 
2269	&4575	&2.40	&-0.02	&1.20		 &-0.11	&0.07	&M  \\
2291	&4746	&2.50	&0	&1.20		 &-0.13	&0.07	&M  \\
2309	&4746	&2.50	&0	&1.20		 &-0.11	&0.04	&M  \\
2311	&4778	&2.50	&-0.02	&1.15		 &-0.06	&0.07	&M  \\
2313	&4674	&2.50	&-0.01	&1.20		 &-0.03	&0.06	&M  \\
 Ruprecht 4 &  &  &  &  &  &    &    \\ 
3	&5180	&2.63	&-0.07	&1.56		 &0.22	&0.10	&M  \\
4	&5150	&2.52	&-0.04	&1.66		 &0.11	&0.12	&M  \\
8	&5190	&2.64	&-0.16	&1.40		 &0.23	&0.02	&M  \\
18	&5040	&3.17	&-0.35	&1.20		 &0.19	&0.07	&NM \\
 Ruprecht 7 &  &  &  &  &  &    &    \\
2	&5160	&2.12	&-0.34	&1.62		 &0.21	&0.00	&M  \\
4	&5105	&2.05	&-0.24	&1.90		 &0.06	&0.07	&M  \\
5	&5230	&2.19	&-0.27	&2.10		 &0.30	&0.00	&M  \\
6	&5230	&2.23	&-0.20	&2.08		 &0.28	&0.11	&M  \\
7	&5150	&2.40	&-0.25	&1.82		 &0.27	&0.00	&M  \\
 NGC 2477  &  &  &  &  &  &    &    \\
4027	&4969	&2.60	&0.10	&1.20    &-0.03	&0.00	&M  \\
4221	&4728	&2.40	&0.19	&1.00   	 &-0.15	&0.04	&M  \\
5043	&5040	&2.60	&0.08	&0.90   	 &-0.29	&0.07	&NM \\ 
5076	&4992	&2.60	&0.18	&1.00   	 &-0.17	&0.01	&M  \\
7266	&4974	&2.60	&0.19	&1.20   	 &-0.21	&0.02	&M  \\
7273	&4993	&2.60	&0.20	&1.15   2	 &-0.23	&0.04	&M  \\
8216	&4945	&2.70	&0.14	&1.20   	 &-0.25	&0.07	&NM \\
 NGC 2506  &  &  &  &  &  &    &    \\
1112	&4969	&2.60	&-0.22	&1.20		 &0.10	&0.20	&M  \\
1229	&4728	&2.40	&-0.22	&1.00		 &-0.09	&0.04	&M  \\
2109	&5040	&2.60	&-0.22	&0.90  	 &0.13	&0.07	&NM \\
2380	&4992	&2.60	&-0.19	&1.00   	 &0.04	&0.35	&M  \\
3231	&4974	&2.60	&-0.22	&1.20   	 &0.11	&0.11	&M  \\
5271	&4993	&2.60	&-0.24	&1.15   	 &0.02	&0.09	&M  \\
 NGC 5822  &  &  &  &  &  &    &    \\
 13292	&5010	&2.8	&0.04	&1.20   	 &-0.05	&0.11	&M  \\
     16450	&4972	&2.6	&-0.02	&1.20   	 &-0.04	&0.11	&NM \\
     18897	&5030	&2.7	&-0.02	&1.00   	 &-0.10	&0.19	&M  \\
      2397	&5036	&2.8	&0.02	&1.10   	 &-0.14	&0.13	&M  \\
NGC 6192  &  &  &  &  &  &    &    \\
 9	&5050	&2.30	&0.19	&1.75   	 &-0.07	&0.07	&M  \\
  45	&5020	&2.55	&0.08	&1.60   	 &0.07	&0.01	&M  \\
   96	&5050	&2.30	&0.13	&2.10  	 &0.04	&0.07	&M  \\
  137	&4670	&2.10	&0.07	&1.80   	 &-0.09	&0.03	&M  \\
NGC 6404  &  &  &  &  &  &    &    \\
5	&5000	&1.00	&0.05	&2.60		 &-0.08	&0.14	&M  \\
 NGC 6583 &  &  &  &  &  &    &    \\
46	&5100	&2.95	&0.40	&1.45		 &-0.17	&0.11	&M  \\
\hline
%\end{tabular}
%\end{table*}
\end{longtable}

\section{Determination of the Mo abundance}

\par
\begin{figure}
\begin{tabular}{cc}
\includegraphics[width=7.5cm]{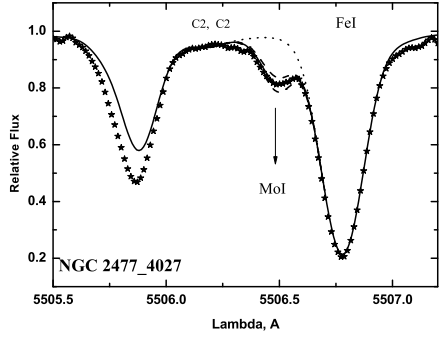}\\
\includegraphics[width=7.5cm]{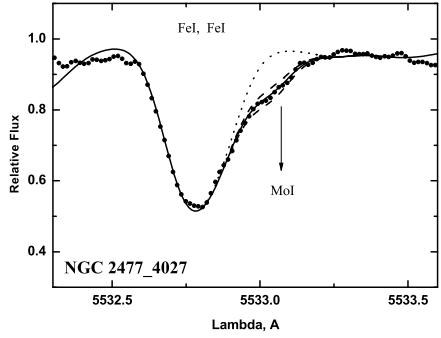}\\
\end{tabular}
\caption{Spectra in the region of Mo I lines at 5506 \AA~ and 5533 \AA~ for NGC 2477 (4027). Observed (asterisks) and calculated (dashed line) depicting a change of 0.1 dex in the Mo abundance and a dotted line which does not factor in the Mo contribution.}
\label{prof}
\end{figure}

The abundances of molybdenum were derived in the LTE approximation from the Mo I lines at 5506.493 \AA~ (log gf = 0.06) and 5533.031 \AA~ (log gf = -0.069) using the Castelli and Kurucz models [56] and a modified spectral synthesis code STARSP [57]. The lists of lines and oscillator strengths were taken from the latest version of the Vienna Atomic Line Database (VALD) [58] revised in 2016. The adopted LTE solar Mo abundance was logA $(Mo)_{\odot}$ = 1.88$\pm$0.08 [59].  
Fig.\ref{prof} illustrates the observed spectrum of star  NGC 2477 (4027) fitted with the calculated (synthetic) spectrum in the region of the Mo lines.  
The derived molybdenum to iron abundance ratios in the open cluster stars relative to the solar [Mo/Fe] ratios are given in Tabl.\ref{parmo}.

\subsection{The estimated impact of the parameter accuracy on the Mo abundance determination}

Tabl.\ref{pardel} presents the estimated impact of the parameter accuracy on the Mo abundance determination in a star of the open cluster Cl* Ruprecht 7 CGM 4	with the atmospheric parameters \Teff/\logg/\Vt/[Fe/H] = 5190/2.64/1.4/-0.16 for the following parameter variation values: $\delta$\Teff = +100 K; $\delta$\logg = +0.2 km/s; $\delta$\Vt = 0.2 km/s, and the fitting of spectrum is +0.03. The total error is given in the last column labeled Total. The parameter variation values taken from an earlier study conducted by Mishenina et al. [43] correspond to the accuracy of the measured parameters.
As can be seen from Tabl.\ref{pardel}, the error in the Mo abundance determination is 0.14 dex. 

\subsection{Comparison with the results obtained by other authors for the open cluster stars}

The abundances of Mo for the clusters in common with this study have only been reported in [60] just for one open cluster, namely Collinder 261. In the study [60], the Mo I lines used to determine the Mo abundances were different from those employed in our study -- in particular, they used the Mo I lines at 5570.44 \AA~ (loggf = -0.34), 5751.41 \AA~ (loggf = -1.01) and 6030.64 \AA~ (loggf = -0.52) with the Mo atomic data taken from [61] and modelled each feature as a single line.
There are no examined stars in the Collinder 261 cluster in common with the study of Overbeek et al. [60]; we can compare only the cluster mean metallicity $<[Fe/H]>$ and Mo abundance $<[Mo/Fe]>$, which are as follows: 
$<[Fe/H]>_{our}$  = -0.01,   $<[Mo/Fe]>_{our}$ = -0.09 and $<[Fe/H]>_{Over}$ = -0.06,  $<[Mo/Fe]>_{Over}$ = -0.05 
%<[Fe/H]>_our = -0.01, <[Mo/Fe]>_our = -0.09 and <[Fe/H]>_Over = -0.06, <[Mo/Fe]>_Over = -0.05 
(our determinations and those reported in [60], respectively). The cluster mean metallicity $<[Fe/H]>_{Over}$ is 0.05 lower while the mean Mo abundance $<[Mo/Fe]>_{Over}$  is 0.04 higher than our values, but they all agree within the errors. 

\begin{table}
\caption{Impact of the parameter accuracy on the Mo abundance determination, OC's star  Ruprecht 7 (4)	(5190/2.64/1.4/-0.16)   }
\label{pardel}
\begin{tabular}{lccccl}
\hline
Element & $\delta$\Teff & $\delta$\logg& $\delta$\Vt & fit &Total	\\
\hline
Mo I & 0.13 &0.06 &0.01 &0.03 &0.14	\\
\hline
\end{tabular}
\end{table}

\section{Results and discussion}

\begin{figure}
\begin{tabular}{c}
\includegraphics[width=7.5cm]{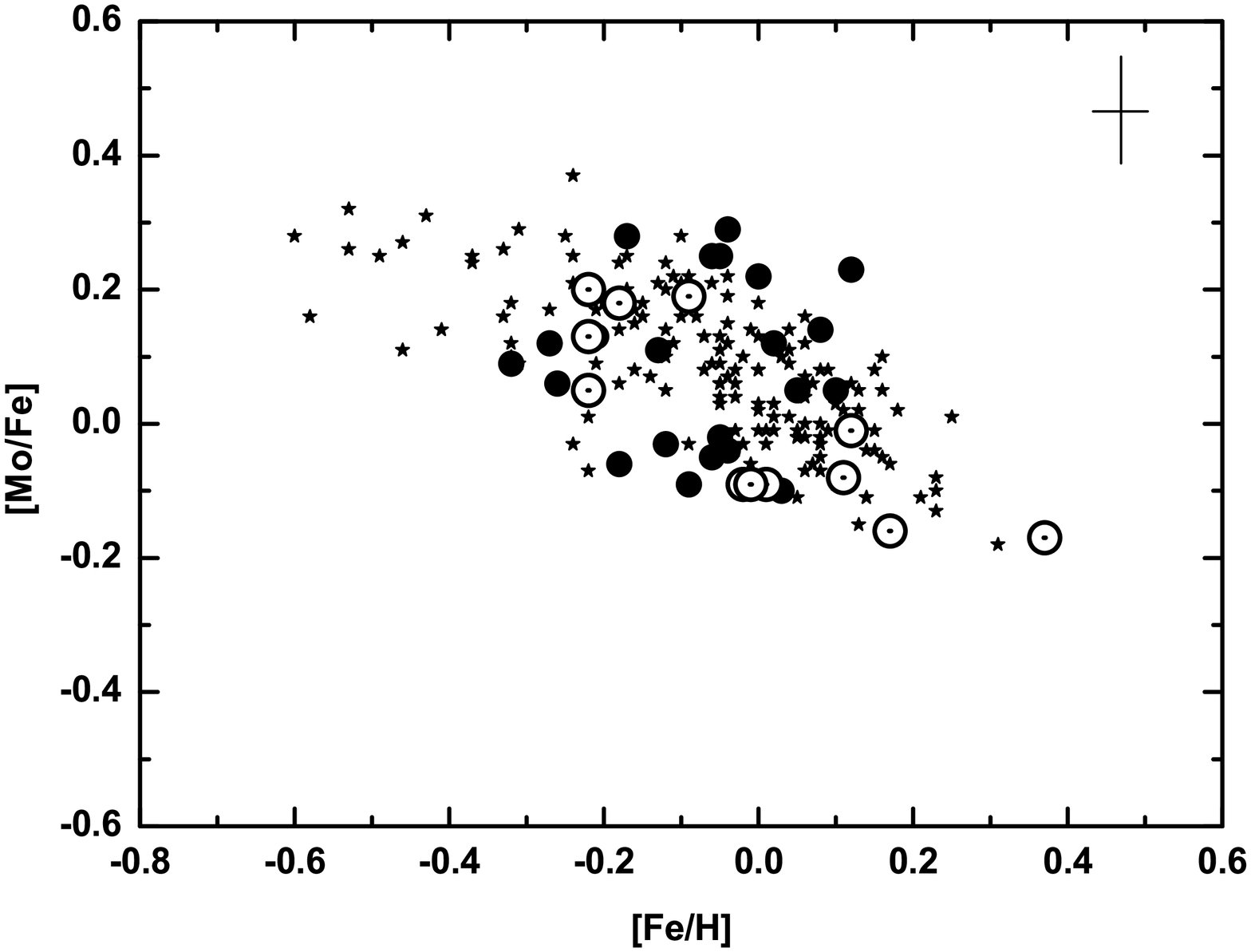}\\
\end{tabular}
\caption{Comparison of the resulting mean Mo abundances for the open clusters (the values derived in this study are plotted using circles with dots inside while those from [60] are marked as filled circles) and for the disc stars [3] (marked as asterisks).}
\label{comp1}
\end{figure}

The mean Mo abundances for the members of the target clusters are given in Tabl.\ref{momean}, 
averaging was carried 
out over the values of the Mo abundance for cluster members (n). 
The 
comparison of the mean Mo abundances for the open clusters in this work and in [60], and also for the disc stars [3] is presented in Fig.\ref{comp1}.

\begin{table}
\caption{The mean metallicities and Mo abundances for the examined clusters.}
\label{momean}
\begin{tabular}{lcccc}
\hline 
Name& $<[Fe/H]>$ & $<[Mo/Fe]>$ & $\sigma$ & n\\
\hline
Berkeley 25& -0.18& 0.18 & 0.08 & 2 \\
Berkeley 73& -0.22& 0.13 & 0.06 &  2\\
Berkeley 75&-0.22& 0.13& -- & 1 \\
Collinder 110&	-0.02& -0.09& 0.08 & 6\\
Collinder 261 &-0.01& -0.09 & 0.04& 5\\
Ruprecht 4 &-0.09& 0.19 &  0.05& 3\\
Ruprecht 7 &-0.22& 0.22 & 0.10& 5 \\
NGC 2477 &0.17 &-0.16 &  0.08 & 5 \\
NGC 2506 &-0.22 &0.04 &  0.08 & 5\\
NGC 5822 &0.01& -0.09 &  0.05 & 3\\
NGC 6192 &0.12 &-0.01 &  0.08 & 4 \\
NGC 6404 &0.11& -0.08 &  -- &  1\\
NGC 6583 &0.4&	-0.17 &  -- &  1 \\
\hline
\end{tabular}
\end{table}

As can be seen in Fig. \ref{comp1}, there are superimposing data sets which are coincident within the errors and show similar dependences (trends) of decreasing Mo abundances with increasing metallicities in the open clusters and disc stars. Moreover, the observational data from [60] show greater number of clusters with the Mo abundances that are high, but still within the range of values for the Galactic disc stars nevertheless.

\subsection{The mean Mo abundances and respective scatter in different stellar populations of the Galactic disc}
 
We compared the mean molybdenum $<[Mo/Fe]>$ and europium $<[Eu/Fe]>$ abundance ratios, as well as respective standard error ($\sigma$), determined for the open cluster stars in this study and in [60], with the mean values for the disc dwarfs from [3] and the open cluster $<[Eu/Fe]>$ ratios reported in a series of papers by Reddy et al. [54, 62, 63, 64]. In Tabl.\ref{mocomp}, europium is selected additionally as the second (reference) element in common with our precedent study [43], based on the same spectra and used the same parameters as in this work, and in [60]. Unfortunately, no Mo abundances were reported in the studies by Reddy et al. [54, 62, 63, 64]. Values n represent the number of either clusters or disc dwarfs.

%Table 5. The mean [Mo/Fe] and [Eu/Fe] ratios and respective standard errors sigma in the examined objects. 

\begin{table*}
\caption{The mean [Mo/Fe] and [Eu/Fe] ratios and respective standard errors ($\sigma$) in the examined objects. }
\label{mocomp}
\begin{tabular}{lcccccccc}
\hline 
Objects/Reference & $<$[Mo/Fe]$>$ & $\sigma$, &$\pm$&   n & $<$[Eu/Fe]$>$ & $\sigma$, & $\pm$ &  n \\
\hline
Galactic disc [3] &   0.08&     0.11 &    163&       0.07 &     0.11&        197  \\
OC's (this work), [43]   &    0.02 &    0.14 &     13 &      0.14  &     0.13 &   11   \\
OC's [60]  &    0.11&    0.14&   23  & -0.05&      0.14&         23  \\
OC's   [54, 62, 63, 64]&  ---   &  --   &    --  &        0.11 &      0.08 &      33  \\
\hline
\end{tabular}
\end{table*}

As can be seen from Tabl.\ref{mocomp}, the mean ratio $<[Mo/Fe]>_{Over}$ = 0.11 is close to the mean value for the disc stars (0.08); however, for two set of OC's the ratios $<[Mo/Fe]>_{our}$ = 0.02 and  
$<[Mo/Fe]>_{Over}$ differ from each other, and this difference makes -0.09. A direct comparison of the results for the open cluster Collinder 261 also yields such a difference (offset). Although this difference is within the Mo abundance determination error (as specified above, this error is about 0.14 dex), it is indicative of the systematic offset which may be associated with different approaches and methods applied in these two studies, in particular different Mo lines, i.e. Mo atomic data used in the present study and in [60].

As for the scatter (dispersion) of Mo abundance, the dispersion for stars inside the considered clusters does not exceed 0.08 dex with the exception of Ruprecht 7 (0.1 dex, see Tabl.\ref{momean} ); at the same time, for the two sets of considered open clusters, the scatter is the same, obtained in this work and in [60], and equals to 0.14 dex, which is slightly larger than for disk stars (0.11 dex), but this is within the determining error of the molybdenum abundance (0.14 dex).

A comparison between the mean Eu abundances obtained in this study $<[Eu/Fe]>_{our}$ and in [60] $<[Eu/Fe]>_{Over}$ shows a difference of about 0.2 dex, the $<[Eu/Fe]>_{Over}$ ratio is lower than our ratios for both the open cluster stars and disc stars (Tabl.\ref{mocomp}). Therefore, we make an additional comparison between the $<[Eu/Fe]>_{our}$ ratio obtained in this study and the $<[Eu/Fe]>_{Reddy}$ ratio reported in a series of studies by Reddy et al. covering a larger sample of clusters [54, 62, 63, 64]. As noted earlier, due to the fact that no Mo abundances are available from [54, 62, 63, 64], we add another comparison (Tabl.\ref{occomp}) with the yttrium and lanthanum abundances reported in [54, 62, 63, 64]. As can be seen in Tabl.\ref{mocomp}, the Eu abundances in the disc stars and open clusters (for the sets sampled in our studies and in [54, 62, 63, 64]) are similar while the values reported in [60] are different from all the others; the Eu abundance scatter in the open clusters (sampled in this study and in [60]) is greater than the cluster values in [54, 62, 63, 64] and those for the disc stars in [3].

%Table 6. Comparison of the [Y/Fe], [La/Fe] and [Eu/Fe] ratios obtained in this study and in a series of the open cluster studies by Reddy et al. for 11 and 33 sampled clusters, respectively.

\begin{table*}
\caption{Comparison of the mean values of [Y/Fe], [La/Fe] and [Eu/Fe] ratios obtained in this study and in a series of the open cluster studies by Reddy et al. [54, 62, 63, 64] for 11 and 33 sampled clusters, respectively }
\label{occomp}
\begin{tabular}{lcccccc}
\hline 
$<[Y/Fe]>$ &    $\sigma$, $\pm$ &  $<[La/Fe]>$ &  $\sigma$, $\pm$ &  $<[Eu/Fe]>$  &   $\sigma$, $\pm$ &   References\\
\hline
0.04   &           0.11 &            0.11  &      0.09     &       0.14 &           -0.13  &      this study\\
0.07  &            0.06  &           0.06 &       0.12    &         0.11 &            0.08  &  [54, 62, 63, 64] \\
\hline
\end{tabular}
\end{table*}
         
The difference resulted from the comparison of the Eu abundances obtained in this study ($<[Eu/Fe]>_{our}$ = 0.14) and those in [60] ($<[Eu/Fe]>_{Over}$  = -0.05) may be associated with a greater number of distant clusters examined in [60] as compared to this study; moreover, these distant clusters may be responsible for some variations in the mean ratios due to the possible presence of the abundance gradients in the Galaxy.

\subsection{The abundances of Mo in stellar clusters as a function of the cluster age and distance to the Galactic centre $R_{GC}$ }

\begin{figure}
\begin{tabular}{c}
\includegraphics[width=7.5cm]{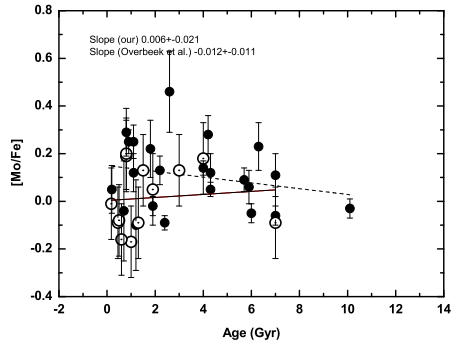}\\
\end{tabular}
\caption{[Mo/Fe] vs. age of clusters is shown for data in this study (empty circles with central dot) and by [60] (filled circles). Slopes from the different sets are shown with continuous line and dashed  line, respectively.}
\label{mofe_aoc}
\end{figure}

\begin{figure}
\begin{tabular}{c}
\includegraphics[width=7.5cm]{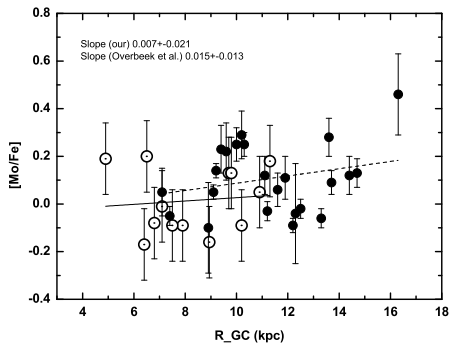}\\
\end{tabular}
\caption{As for Figure 3, but [Mo/Fe] vs. Galactocentric radius $R_{GC}$  is shown. 
}
\label{mofe_roc}
\end{figure}

The Mo abundance behaviour as a function of the cluster age and distance to the Galactic centre $R_{GC}$ was investigated in [60]. However, the reported results were subject to a sufficient degree of uncertainty as the absence of data from other studies did not allow the authors to compare their results and thus draw more reasonable conclusions. Let us examine the Mo abundance behaviour taking into account our newly obtained determinations.
The dependences of the [Mo/Fe] ratios on the cluster age and distance to the Galactic center (i.e. Galactocentric radius) $R_{GC}$ are plotted in Figs. \ref{mofe_aoc}, and \ref{mofe_roc}, respectively, for two sets of clusters examined in this study and in [60].

As is seen from Fig. \ref{mofe_aoc}, there is hardly any correlation between the Mo abundance and age of the cluster. From our set of data, the slope is 0.006$\pm$0.021 dex Gyr$^-1$. From [60] the slope is -0.0121$\pm$0.013 dex Gyr$^-1$, that is consistent with our results within the errors. As reported in [60], “there is no visual suggestion of a trend with age when the cluster Berkeley 17 (which is 10.2 Gyr old) is excluded from the sample”. Our results  confirm this conclusion.

As regards the dependence of [Mo/Fe] on the Galactocentric radius $R_{GC}$  (Fig. \ref{mofe_roc}), our results show a small gradient of 0.007$\pm$0.020 dex kpc$^-1$. The slope for the data reported by Overbeek et al.[60] is 0.014$\pm$0.013 dex kpc$^-1$. Also in this case, our results are consistent with [60], confirming that there is no significant trend with Galactocentric radius. 

We also considered the abundance trend of the combined Mo data with respect to  the age and Galactocentric radii for the entire list of clusters in the combined sample. There is no correlation between the Mo abundance ratio [Mo/Fe] and age of the cluster (slope $=$ -0.001$\pm$0.010 dex Gyr-1) for the combined sample of clusters. As noted earlier, Mo is made  by  different processes consistently with this scenario, the resulting  Mo abundance trend  does not follow the trend of typical s-process elements like Zr, La and Ba [65]. This confirms  that it is not the s - process that dominates the Mo production in the galactic disk (e.g., [1, 6, 3]).

The compiled [Mo/Fe] trend with the Galactocentric radius $R_{GC}$ for all the sampled clusters is represented with a slope of 0.018$\pm$0.009 dex kpc$^-1$, which is similar to the slope of 0.015 $\pm$ 0.013 dex kpc$^-1$ reported in [60], but is of a higher statistical significance level (p-value). It is essential and interesting to compare the degree of correspondence between this trend for Mo and a compiled trend for Eu, which is an element produced mainly by the r-process, reported in [60] (a slope of 0.039 $\pm$ 0.010 dex kpc$^-1$) and other studies. Jacobson \& Friel [65] reported a slope of 0.047 dex kpc$^-1$ while Yong et al. [66] determined the abundance gradient for two ranges of distances, in particular 0.07 $\pm$ 0.01 dex kpc$^-1$ for $R_{GC}$   $<$ 13 kpc and 0.01 $\pm$0.00 dex kpc$^-1$ for $R_{GC}$  $>$ 13 kpc. As discussed in Overbeek et al. [60], a scatter in the [Eu/Fe] ratios among different sources was up to 0.3 dex due to the differences in atomic data and techniques used for the EW measurements of lines. Our data are in agreement with such a value: the difference between the mean abundances for the clusters examined in this study and in [60] is 0.19 dex (Table 5). However, when considering only Eu abundance in clusters sampled in [66, 67] within the range of Galactocentric radii adopted in [60], Overbeek et al.  [60] found a slope of [Eu/Fe] of 0.031 $\pm$ 0.023 dex kpc$^-1$, which was consistent with the resulting slope of the regression line of 0.039$\pm$0.010 dex kpc$^-1$ [60]. Despite some inconsistency in the determinations of the slope for [Eu/Fe] by different authors, the slope of the Mo abundance with the Galactocentric radius derived in this study is significantly smaller compared to that of Eu. If we consider the [Mo/Fe] ratios for the target clusters within the range of $R_{GC}$ $<$ 13 kpc, the relevant slope will be 0.007$\pm$0.012 dex kpc$^-1$. Since Mo does not behave like the r-process element Eu, this also confirms that the production of Mo is not dominated by the the r-process. However, it is not feasible to carry out a quantitative estimation of these contributions.
 
\begin{figure}
\begin{tabular}{c}
\includegraphics[width=7.5cm]{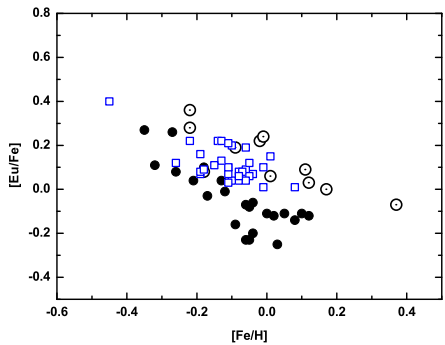}\\
\end{tabular}
\caption{[Eu/Fe] vs. [Fe/H] for three samples of open clusters: the values from [43] are plotted using circles with dots inside; the data from [60] are marked as filled circles while those from  [54, 62, 63, 64] are represented with blue squares.
}
\label{eufe_oc}
\end{figure}

For the comparison of the Mo and Eu  trends, we used the Eu abundance determinations for open clusters from [43], [60] and [54, 62, 63, 64], as well as those for the disc stars reported in [3]. Note that in all the listed references the Eu abundances were determined using the same 6645 \AA~ line, with atomic data from [68] and by a calculated synthetic spectrum as the line is of hyperfine structure and is slightly blended, though not as much as the other Eu lines [60]. 
Fig. \ref{eufe_oc} illustrates a comparison between the Eu abundances determined by different authors for three samples of open clusters [43, 60, 54, 62, 63, 64].

\begin{figure}
\begin{tabular}{c}
\includegraphics[width=7.5cm]{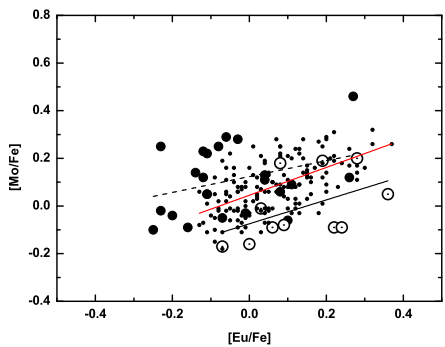}\\
\end{tabular}
\caption{[Mo/Fe] vs. [Eu/Fe] for the Galactic disc sample from [3] (dots) and for the open cluster samples from [43] (circles with dots inside) and [60] (filled circles).
}
\label{moeu}
\end{figure}

A common trend of [Eu/Fe] with [Fe/H], i.e. decreasing Eu abundance with increasing metallicity, is obvious for all three samples, but the slope for the data from [60] is steeper. A correlation between [Mo/Fe] and [Eu/Fe] is shown in Fig. \ref{moeu}; the slope is 0.583 $\pm$ 0.068 for the disc stars from [3] (red line); 0.501 $\pm$ 0.305 for the open clusters from [43] (black line) and 0.333 $\pm$ 0.207 for the sample from [60] (dashed line).

When neglecting the difference between systematic errors in these studies, it is evident that all three samples exhibit similar trends of [Mo/Fe] with [Eu/Fe]. The resulting slopes confirm that the contribution of the r-process to the molybdenum enrichment is not significant as compared to its contribution to the europium abundance.  

\section{Summary and conclusions}

1.	The abundances of molybdenum have been derived for the stars in 13 open clusters in the first time. 

2.	Similar abundance trends of decreasing Mo abundances with increasing metallicities have been found for the open clusters and disc stars; such a behaviour pattern suggests a common origin of the examined populations.

3.	According to the determinations in this study, as well as in the literature, the open cluster stars exhibit a larger scatter (dispersion) of [Mo/Fe] values as compared to the disc dwarfs (is slightly greater, 0.14 dex versus 0.11 dex). It may be due to both methodological errors in the abundance determinations and actual scatter associated with inhomogeneities in the disc enrichment at different distances to the Galactic center.

4.	The results of this study for the open clusters confirm the absence of OCs age-dependence of the Mo abundance, as reported earlier in [60].

5.	The dependence of the Mo abundance on the Galactocentric radius found in this study suggests that the s-process (r-process) contribution to the Mo enrichment is significantly smaller than its contribution to typical s-process (r-process) elements. As a diagnostic, we used Zr, Ba and La as s-process elements, and Eu as r-process element.

6.	An analysis of the obtained results and the dependences of the Mo content on the age of clusters and distances from the center of the Galaxy manifest that there is no correlation between the Mo abundance ratio [Mo/Fe] and age of the cluster (slope = -0.001$\pm$0.010 dex Gyr$^-1$) for the combined sample of clusters. The compiled [Mo/Fe] trend with the Galactocentric radius $R_{GC}$ for all the sampled clusters is represented with a slope of 0.018$\pm$0.009 dex kpc$^-1$, it is of a higher statistical significance in comparison with earlier obtained data. This slope is different from those obtained for s- or r-processes n-capture elements. 

\section*{Acknowlegements}
M.P. thanks the support support to NuGrid from STFC (through the University of Hull's Consolidated Grant ST/R000840/1), and access to {\sc viper}, the University of Hull High Performance Computing Facility. MP also acknowledges the support from the "Lendulet-2014" Program of the Hungarian Academy of Sciences (Hungary), from the ERC Consolidator Grant (Hungary) funding scheme (Project RADIOSTAR, G.A. n. 724560), by the National Science Foundation (NSF, USA) under grant No. PHY-1430152 (JINA Center for the Evolution of the Elements). This article is based upon work from the ChETEC COST Action (CA16117), supported by COST (European Cooperation in Science and Technology).
%\end{document}

\end{document}